\documentstyle[preprint,prd,aps,12pt]{revtex}
\preprint{
$
\begin{array}{r}
\text{hep-ph/9408403} \\ \text{LAVAL-PHY-11-94}
\end{array}
$
}
\tighten

\begin{document}
\author{B. Dion, L. Marleau and G. Simon}
\address{D\'epartement de Physique, Universit\'e Laval\\
Qu\'ebec, Canada, G1K 7P4}
\title{Skyrmions from a Born-Infeld Action }
\date{July 1994 }
\maketitle
\draft

\begin{abstract}
We consider a geometrically motivated Skyrme model based on a general
covariant kinetic term proposed originally by Born and Infeld. We introduce
this new term by generalizing the Born-Infeld action to a non-abelian $SU(2)$
gauge theory and by using the hidden gauge symmetry formalism. The static
properties of the Skyrmion are then analyzed and compared with other
Skyrme-like models.
\end{abstract}


\section{Introduction}

For many years, the non-linear $\sigma $-model has been known to provide an
approximate description of hadronic physics in the low-energy limit.
Moreover, its finite-energy configuration space exhibits a non-trivial
topological structure which allows the existence of static and finite field
configurations other than the trivial vacuum. Unfortunately, the $\sigma $%
-model Lagrangian being of second order in field derivatives, these
so-called soliton configurations are not energically stable in ordinary
three-dimensional space.

In a celebrated series of articles, Skyrme \cite{skyrme} proposed to modify
the Lagrangian in order to include a quartic term that prevents the model
from being renormalizable. His model provides the ground state properties of
the nucleon to within about 30\% accuracy. The new term insures dynamically
stable solitons (or Skyrmions), which Skyrme suggested to identify with
baryons. Little physical significance could be attached to this term
however, except for the fact that it could arise as a manifestation of
higher-spin meson exchanges.

In recent years, renewed interest has been paid to the Skyrme model in the
context of weak interactions as well as in low-energy QCD with the
introduction the hidden gauge symmetry formalism. It was then understood
that the Skyrme quartic term could be formally interpreted as the kinetic
contribution of a vector gauge field in the limit where the mass of this
field becomes infinite. Indeed, various authors \cite{bando} have shown that
the nonlinear $SU(2)$ $\sigma $-model with spontaneously broken chiral
symmetry was equivalent to a linear model based on the group $SU(2)_L\otimes
$ $SU(2)_R$ $\otimes $ $SU(2)_V$ where $SU(2)_V$ is assumed to be a hidden
symmetry. This new vector field becomes dynamical when we add a kinetic term
of the form $Tr\left[ F_{\mu \nu }F^{\mu \nu }\right] $ to the Lagrangian.
When the mass of the vector field is taken to be infinite, the latter
decouples and we find \cite{dobado} $F_{\mu \nu }\rightarrow \left[ L_\mu
,L_\nu \right] $ where $L_\mu =U^{\dagger }\partial _\mu $ $U$ and $U$ is
the pion field. We then recover the Skyrme Lagrangian.

On the other hand, the requirement of gauge invariance is not sufficient to
fix a precise form of the gauge term. It could in principle include -- in
addition to a usual gauge kinetic term -- terms with higher orders, or even
non-polynomial functionals, of the field derivatives. Of course, in these
cases, the resulting theory loses explicit renormalizability but it is hoped
that there exists some mechanism of cancellation of divergences that may
render physical results finite. In a paper dating back to 1934, Born and
Infeld \cite{born} proposed a general covariant action function, which has
the attractive features of arising directly from the metric. In this work,
we will consider a generalization of the Born-Infeld action to describe
chiral dynamics. Indeed, there exist a number of analogies between chiral
dynamics and general relativity \cite{pak} and the idea of a direct
connection is very appealing. They are both intrinsically non-linear and are
characterized by a non-trivial geometrical structure. Moreover, the
compactification of space from $R^3$ to $S^3$- due to the necessary boundary
conditions - implies that the soliton field effectively ``sees'' a closed
physical space $S^3$. Thus the analogy with a closed Einstein universe with
hadronic size radius.

The Born-Infeld action was originally built as a revival of the old model of
the electromagnetic origin of mass. It has the property of insuring absolute
finiteness of energy and of reducing to the usual $Tr\left[ F_{\mu \nu
}F^{\mu \nu }\right] $ Maxwell form in the low-energy limit. Another
interesting feature of this theory is the non-linear nature of the field
equations. The latter are distinguished by a mass scale $M$ which suggests
an analogy with the notion of effective Lagrangians where we have integrated
over the heavy particles' degrees of freedom. Whereas Born and Infeld
introduced this action in the context of $U(1)_{EM}$, we intend in this work
to generalize it to $SU(2)$ chiral dynamics, using the hidden gauge symmetry
formalism mentioned above. Our analysis and computations mainly deal with
hadron physics which is the best prototype available (since physical
quantities are easier to identify) but the idea and most conclusions could
easily be transposed to other contexts, e.g. weak Skyrmions. The Skyrmions
we described are themselves characterized by a length scale $1/M$.

The remainder of this paper is divided in four sections. First, we present
the concept and principal features of the Born-Infeld action and introduce a
proper choice of generalization to $SU(2)$. Then, we build an effective
chiral Lagrangian using the HGS formalism. We finally proceed to derive and
discuss the physical properties of resulting Skyrmions in the context of
hadron physics.

\section{The Born-Infeld Action}

The original purpose of Born and Infeld \cite{born} was to introduce a new
electrodynamic field theory in which the self-energy of a point charge would
be finite. They proposed to apply the principle of finiteness - which says
that all physical quantities should be finite everywhere - to the
electromagnetic field by postulating an absolute value $b$ such that $F_{\mu
\nu }\rightarrow b$ in the high-energy limit. The $b$ constant has
dimensions $\left[ M\right] ^2$ where $M$ can be interpreted as a scale
parameter. The new kinetic term they derived bears a striking resemblance to
the relativistic expression of the kinetic energy of a particle, with the
velocities $v$ and $c$ replaced by the field strengths $F_{\mu \nu }$ and $b$%
{}.

Apart from the finiteness of energy, our interest in the Born-Infeld
Lagrangian is enhanced by the following properties:

({\it i}) An intrinsic length scale reminiscent of an effective Lagrangian
with the heavy particles' degrees of freedom integrated over.

({\it ii}) The Lagrangian is constructed out of the two Maxwell invariants
of only $F_{\mu \nu }F^{\mu \nu }$ and $F_{\mu \nu }^{*}F^{\mu \nu }$, not
derivatives such as $[D^\lambda F^{\mu \nu }]^2$.

({\it iii}) It is intrinsically non-linear but still, it can be quantized
\cite{dirac} and is the only causal spin-1 theory \cite{plebanski} aside
from the gauge Lagrangian, $-\frac 14F_{\mu \nu }F^{\mu \nu }.$

({\it v}) Most of all, it is geometric by nature. It is one of the simplest
non-polynomial Lagrangians that is invariant under the general coordinate
transformations.

The Born-Infeld action is based on the invariant measure written as%
$$
\delta \int {\cal L}d\tau =0,\qquad (d\tau =dx^0dx^1dx^2dx^3)
$$
where the metric is considered to be represented by a general
non-symmetrical tensor $a_{\mu \upsilon }$. The symmetrical part is the
usual space-time metric $g_{\mu \nu }$ and the anti-symmetrical part is
identified with the electromagnetic field $F_{\mu \nu }.$ The general
covariant action which reproduces the Maxwell action in the limit of flat
space-time and lowest order in the field strength is then

\begin{equation}
\label{lbi1}{\cal L}_{BI}=b^2\ \left( \sqrt{-\left| g_{\mu \nu }+F_{\mu \nu
}\right| \ }-\sqrt{-\left| g_{\mu \nu }\right| }\right)
\end{equation}
where $\left| g_{\mu \nu }\right| =\det g_{\mu \nu }.$

Assuming flat space-time (i.e. $g_{\mu \nu }={\rm diag}(1,-1,-1,-1)$), the
last expression can be written in the form

\begin{equation}
\label{lbi1.5}{\cal L}_{BI}=b^2\left( \sqrt{1+K-G^2}-1\right)
\end{equation}
where
\begin{eqnarray}
   K    & = & \frac 1{b^2}F_{\mu \nu }F^{\mu \nu } \\
   G   & = & \frac 1{2b^2}F_{\mu \nu }F^{*\mu \nu }
\end{eqnarray} and $F^{*\mu \nu }\ $is the dual field of $F^{\mu \nu }$. In
the weak field approximation $(\left| F\right| \gg \left| G^2\right| )$ or
in the case of static configuration $(G=0)$ we can write

\begin{equation}
\label{lbi2}{\cal L}_{BI}=b^2\left( \sqrt{1+\frac 1{b^2}%
\mathop{\rm Tr}
\left[ F_{\mu \nu }F^{\mu \nu }\right] }-1\right)
\end{equation}

In the low-energy limit, $F_{\mu \nu }\rightarrow 0$ and one recovers the
familiar Maxwell form which in our notation is

\begin{equation}
\label{lbi3}{\cal L}_{BI}\rightarrow \frac 12%
\mathop{\rm Tr}
\left[ F_{\mu \nu }F^{\mu \nu }\right]
\end{equation}

It is then easy to derive the conservation laws of energy-momentum and to
describe the finite-energy electron by a static solution with spherical
symmetry and finite spatial extension. One finds that the $\left| \frac 1{b^2%
}Tr\left[ F_{\mu \nu }F^{\mu \nu }\right] \right| $ term to tend
asymptotically to 1 in the high energy limit.

In order to describe $SU(2)$ hadron physics, we need to generalize the
Born-Infeld idea to non-abelian $SU(2)$ gauge theory. This is done by
introducing a non-abelian metric:%
$$
g_{\mu \nu }+F_{\mu \nu }\rightarrow {\bf a}_{\mu \nu }\equiv g_{\mu \nu }%
{\bf 1}+F_{\mu \nu }^a\tau ^a
$$
where $\tau ^a$ are the Pauli matrices. The Lorentz structure of the metric
is similar to $U(1)$ case so the BI action will again involve the
determinant of the metric (over space-time indices)
\begin{equation}
\label{lbi4}\left| {\bf a}_{\mu \nu }\right| =\frac 1{4!}\epsilon ^{\mu \nu
\rho \sigma }\epsilon ^{\alpha \beta \gamma \delta }{\bf a}_{\mu \alpha }%
{\bf a}_{\nu \beta }{\bf a}_{\rho \gamma }{\bf a}_{\sigma \delta }
\end{equation}
But the Lagrangian must be an $SU(2)$ singlet which involves an operation on
the $SU(2)$ component of $\left| {\bf a}_{\mu \nu }\right| .$ The simplest
extension is given by taking the trace \cite{hagiwara}. The Lagrangian then
takes the form:
\begin{equation}
\label{lbi5}{\cal L}_{BI}^{SU(2)}=\frac{b^2}2\ \left( \sqrt{-%
\mathop{\rm Tr}
\left| {\bf a}_{\mu \nu }\right| \ }-\sqrt{-%
\mathop{\rm Tr}
\left| g_{\mu \nu }{\bf 1}\right| }\right) =b^2\left( \sqrt{1+K-\frac 13%
(G^2+2G_S^2)}-1\right)
\end{equation}
where
\begin{eqnarray}
   K    & = & \frac 1{b^2}F^a_{\mu \nu }F^{a\mu \nu } \\
   G   & = & \frac 1{2b^2}F^a_{\mu \nu }F^{*a\mu \nu }\\
   G_S   & = & \frac 1{4b^2} [ F^a_{\mu \nu }F^{*b\mu \nu } + F^b_{\mu \nu
}F^{*a\mu \nu } ].
\end{eqnarray}  However, in general, any non-polynomial Lagrangian is
expected to be a function of nine gauge invariants \cite{roskies}. Some of
these invariants will not contribute because of the symmetries in the
definition of the determinant in (\ref{lbi4}) but otherwise one can think of
a very special case where the operator ${\cal O}$ is such that%
$$
{\cal O(}\left| {\bf a}_{\mu \nu }\right| )\ =-(1+\frac 12K)^2,\quad {\cal O(%
}\left| g_{\mu \nu }{\bf 1}\right| )=-1
$$
and this leads to the usual gauge kinetic term ${\cal L}_{BI}^O=\frac 12%
\mathop{\rm Tr}
F_{\mu \nu }^aF^{a\mu \nu }$ which has its origin in the metric but no
longer constrains the fields to be finite.

In this work, we will only consider the simplest form of non-abelian
generalization of the Born-Infeld action in (\ref{lbi5}) motivated by its
constraint of finiteness on the fields.

We now give a brief review of the concepts behind the hidden gauge symmetry
formulation.

\section{Hidden Gauge Symmetry}

The HGS formalism is based on the manifold $SU(2)_L\otimes SU(2)_R$ $\otimes
$ $SU(2)_V$ where $SU(2)_V$ is gauged. The most general Lagrangian involving
only two field derivatives is expressed as
\begin{equation}
\label{l2}{\cal L}_2=-\frac{f_\pi ^2}4Tr\left( L^{\dagger }D_\mu
L-R^{\dagger }D_\mu R\right) ^2-ag^2\frac{f_\pi ^2}4Tr\left( L^{\dagger
}D_\mu L+R^{\dagger }D_\mu R\right) ^2
\end{equation}
where $L(x)\in SU(2)_L$ and $R(x)\in SU(2)_{R\text{ }}$. The covariant
derivative $D_\mu $ reads
\begin{equation}
\label{du}D_\mu =\partial _\mu -igV_\mu ^k\cdot \frac{\tau ^k}2
\end{equation}
where $V_\mu $ stands for the hidden gauge field. The first part of equation
is equivalent to the gauged non-linear $\sigma $-model since it can be
brought to the form
\begin{equation}
\label{lnls}{\cal L}_{NL\sigma }=-\frac{f_\pi ^2}4Tr\left( D_\mu U^{\dagger
}D^\mu U\right)
\end{equation}
where the classical Euler-Lagrange equation corresponding to (\ref{l2}) has
been used for the auxiliary field
\begin{equation}
\label{vu}V_\mu =\frac 1{2ig}\left( L^{\dagger }\partial _\mu L+R^{\dagger
}\partial _\mu R\right) .
\end{equation}

The hidden gauge field becomes dynamical when we add a kinetic piece to the
Lagrangian.. Here, we use the Born-Infeld form and write

\begin{equation}
\label{l}{\cal L}={\cal L}_2+{\cal L}_{BI}^{SU(2)}
\end{equation}
The vector boson $V$ acquires its mass from the same mechanism as the
standard gauge bosons with result, $m_V^2=ag^2f_\pi ^2$. In the $%
m_V^2\rightarrow \infty $ limit, we can show that the field strength $F_{\mu
\nu }\rightarrow \frac 1{4e}Tr\left[ U^{\dagger }\partial _\mu U,U^{\dagger
}\partial _\nu U\right] .$ Writing $L_\mu =U^{\dagger }\partial _\mu U$, we
now have
\begin{equation}
\label{ltot}{\cal L}_{tot}=-\frac{f_\pi ^2}4%
\mathop{\rm Tr}
\left( L_\mu L^\mu \right) +b^2\left( \sqrt{1+\frac 1{16e^2b^2}%
\mathop{\rm Tr}
\left[ L_{\mu ,},L_\nu \right] ^2}-1\right)
\end{equation}

This is the Lagrangian form we will be using in the following section to
derive the Skyrmion phenomenology when applied to hadrons. Note that here,
we could have written down this Lagrangian right from the beginning without
any reference to HGS. The anti-symmetrical part of the metric would be due
to the scalar self-interaction. The idea that the Lagrangian (\ref{ltot})
may come from HGS is however attractive for two reasons. First, it is a
straightforward generalization of the Born-Infeld idea which involved,
originally, gauge fields. Secondly, it provides a natural scale for symmetry
breaking, i.e. the Planck mass. Indeed, the HGS formalism introduces a gauge
symmetry which has the effect of stabilizing the Skyrmion but only when the
large vector mass limit is taken. In other words, the symmetry must be
effectively broken at all but very large scales compared to the Skyrmion
scale. In the above treatment however, the stabilizing term is introduced
through the metric via the Born-Infeld action and so it seems only natural
that a large scale ---of the order of the Planck scale--- could be involved.

In the next section, we look at the phenomenology of the system (\ref{ltot}%
). In priciple it could be used in several contexts: low hadron physics,
weak Skyrmions where the Born-Infeld action could originate from a unified
theory or even qualitons. Here, we will only consider the Skyrmions from the
point of view of hadron physics mainly the context is well defined and it is
easier to compare to the results to that of the original Skyrme model.

\section{Phenomenology}

We use \thinspace the standard semi-classical quantization technique of
Adkins et al. \cite{adkins} in order to derive the Skyrmion phenomenology
based on (\ref{ltot}). We consider spherically symmetric and static Skyrmion
configurations. Using the hedgehog ansatz $U=\exp [i\tau \cdot \widehat{r}%
F(r)]$ where $F$ is the chiral angle, we obtain the static mass
\begin{equation}
\label{ms}M_S=4\pi \left( \frac{f_\pi }e\right) \int r^2dr\left[ \frac{\sin
^2F}{r^2}+\frac 12F\ ^{\prime 2}-c\left( \left( 1-K\right) ^{\frac 12%
}-1\right) \right]
\end{equation}
where we have made the change of variable, $r\rightarrow a_0\ r$ , where $%
a_0=\frac{ef_\pi }{\sqrt{2}}$ , $c=\frac{b^2}{e^2f_\pi ^4}\ $(note that for $%
c\rightarrow \infty $ one recovers the Skyrme model) and where

\begin{equation}
\label{k}K=\frac 1c\frac{\sin ^2F}{r^2}\left( 2F^{\prime 2}+\frac{\sin ^2F}{%
r^2}\right)
\end{equation}
A few comments are in order regarding the classical stability of the
Skyrmion due to the non-polynomial nature of this Lagrangian. First, the
Derrick theorem for testing stability against scale transformation does not
trivially apply here. The argument of the square root in (\ref{ms}) must
remain positive, constraining scale transformation to scales only down to a
critical value. In other words, the constraint of a real static energy
---not finiteness of the static energy like in the Skyrme model--- prevents
the size of the Skyrmion from shrinking to zero. On the other hand, the
static energy is always positive since $-c((1-K)^{\frac 12}-1)>0.$

Minimizing $M_S$ with respect to $F$, we obtain the Euler-Lagrange equation

\begin{eqnarray}
    0  & = &[-2 \sin F \cos F+2 r F'+r^{2} F''] \nonumber \\ & &
+\left( 1-K\right) ^{-\frac{1}{2}} \left[2 \sin F \cos F \left(
F'^{2}-\frac{\sin ^{2}F}{r^{2}} \right)+2 \sin ^{2}F F'' \right] \nonumber \\ &
&
+\frac{1}{c} \left( 1-K\right) ^{-\frac{3}{2}} \left[ 4 \frac{\sin ^{2}F
F'}{r}\left( \left( \frac{F'}{r} \sin F \cos F - \frac{\sin
^{2}F}{r^{2}}\right)  \right.\right. \nonumber \\ & &
\hspace{39 mm}\left. \left. \left( F'^{2}+ \frac{\sin ^{2}F}{r^{2}}
\right)+\frac{F'}{r} \sin ^{2}F F'' \right) \right]
\end{eqnarray}

As in the case of the standard Skyrme equation, {}(20) admits no exact
analytical solution and has to be solved numerically. Under rotation, we
find that the Skyrmion inertia is given by
\begin{equation}
\label{i}{\cal I}=\frac{8\pi }{3e^3f_\pi }\ \int r^2dr\left[ 1+\left(
1-K\right) ^{-\frac 12}\left( \frac{\sin ^2F}{r^2}+F^{\,\prime 2}\right) \
\right]
\end{equation}
In computing the preceding expression we have neglected terms which are
quartic (or higher) in time derivatives. This approximation is only
justified under the assumption that the Skyrmion rotates slowly in the
semi-classical limit.

Similarly, we find the axial coupling constant to be given by

\begin{equation}
\label{ga}g_A=-\frac{8{\pi }}{3e^2}\int r^2dr\left[ \frac 12F^{\prime }+%
\frac{{\rm \sin }F{\rm \cos }F}r\left[ 1+(1-K)^{-\frac 12}\left( \frac{{\rm %
\sin }^2F}{r^2}+F^{\prime }{}^2\right) \right] +(1-K)^{-\frac 12}\frac{{\rm %
\sin }^2F}{r^2}F^{\prime }\right]
\end{equation}

In the numerical calculations we carried out a power series expansion of the
above expressions and we truncated those series to a finite order. The
original Skyrme model corresponds to a truncation at fourth order, or
equivalently to the limit $c=\infty $. The experimental masses of the
nucleon ($M_N=939$ MeV) and of the $\Delta $ ($M_\Delta =1232$ MeV) were
used as input to determine the $f_\pi $ and $e$ parameters. On the other
hand, $c$ was left as a free parameter.

The numerical results were computed:

{\it (i)} For different values of $c$, with the order of truncation fixed at
24 (Table I)

{\it (ii)} For truncation at different orders, with the parameter $c$ fixed
at 1 (Table II).

\section{Discussion and Conclusions}

{}From Tables I and II, we immediately see that our numerical results seem to
converge toward the experimental data in the $c\rightarrow 0$ and
infinite-order limits. As infinite-order truncation corresponds to the exact
solution of (20), it would seem preferable to solve the latter equation
directly, for an arbitrary value of $c$ and without resorting to a power
series expansion. However we have found this alternative to be
impracticable. Indeed, the numerical solution requires a finer tuning of the
boundary conditions ({\it i.e. }the slopes of the profile at $r=0$ and $%
r=\infty $) as higher orders are added. The reason for this behavior is that
the solution for small distance saturates $K$ near $K\cong 1$ for a large
range of values $0\leq r\leq r_0$ in order for the series to converge in
which case, the chiral angle behavior is very close to that of an $\arcsin
(ar)$. This numerical instability has also been noted when the
above-mentioned $c\rightarrow 0$ and infinite-order limits. They have also
been observed in several other infinite-order models and are fairly well
understood \cite{marleau} .

Looking at Tables I and II, we note that the results show very little
variation in the latter limits, and thus our approximation to the exact
solution seems to be working rather well. Although minimal in the case of
some observables (more specifically the magnetic moments and the mean
radii), the improvement achieved by our model with respect to the original
Skyrme model is significant. For instance, the results for $f_\pi $ are in
much better agreement with the experimental value. We obtain a $\sim 10\%$
accuracy with $c=0.01$, as compared to the $\sim 30\%$ accuracy for the
Skyrme model. The results for $g_A$ are also better, although they are still
$\sim 30\%$ away from the experimental value. This was to be expected, as
all previous Skyrme-like models \cite{marleau} showed the same feature. This
gap is most probably not related to the particular choice of model, but
rather to a more fundamental flaw in the approach (quantification technique,
...).

In Fig. 1 and Fig. 2, we show the behavior of the chiral angle $F(r)$ with
the same choice of parameters as in tables I and II, whereas in Fig. 3 and
Fig. 4, we show the behavior of $K(r)$ as defined in (\ref{k}). We see from
Fig. 1 the behavior of the chiral angle to become increasingly smoother in
the low-$c$ limit. On the other hand, Figs. 3 and 4 show that the $K$ term
exhibit a plateau for small radial distances --- the plateau extends further
for small values of $c$. Another characteristic we can extract from Fig. 4
is that the $K\rightarrow 1$ behavior of the solution is only fully expanded
for high orders ($\geq 60$).

To conclude, we have considered the general covariant action suggested by
Born and Infeld as an alternative to the usual stabilizing term in the
context of the Skyrme model. The Skyrmion has therefore the attractive
feature of having a geometric origin. Moreover, although the results we have
obtained show only a slight improvement in general in the description of the
static properties of the nucleons, they are in better agreement with the
experimental data than the original Skyrme model.

\section{Acknowledgments}

The authors would like to thank C. Bachas for many helpful discussions and
for getting us interested in the Born-Infeld action. L.M. is grateful for
the hospitality of the Service de Physique Th\'eorique (CEA-Saclay), where
part of this work was done. This work was supported in part by the Natural
Sciences and Engineering Reasearch Council of Canada, by the Fonds pour la
Formation de Chercheurs et l 'Aide \`a la Recherche du Qu\'ebec and by the
Fondation A.F.D.U. Qu\'ebec (G.S.).

\begin{table}
\caption{Numerical results obtained for some physical observables for different
values of the parameter c, defined in equation (10). The c=$\infty$ corresponds
to the standard Skyrme Lagrangian.}
 \begin{tabular}{cddddddd}
    \centering{$$}  &  {c=$\infty$}  & {c=1}  & {c=0.5 }  & {c=0.1}  & {c=0.05}
 & {c=0.01} & {Expt} \\\hline
    \centering{$f_{\pi}$ (MeV)}         &  64.6   & 76.8   & 78.2   & 80.8
 & 81.6  & 82.8   & 93  \\
    \centering{$10^{3} \epsilon^{2}$}   &  4.22   & 1.63   & 1.31   & 0.74
& 0.57  & 0.29   &     \\
    \centering{$<r^2_{E}>^{1/2}$ (fm)}  &  0.59   & 0.61   & 0.61   & 0.62
 & 0.62  & 0.62   & 0.72 \\
    \centering{$<r^2_{M}>^{1/2}$ (fm)}  &  0.92   & 0.86   & 0.86   & 0.85
 & 0.85  & 0.85   & 0.81 \\
   \centering{$\mu_{p} (\mu_{N})$}      &  1.87   & 1.89   & 1.90   & 1.90
& 1.90  & 1.91   &2.79  \\
   \centering{$\mu_{n} (\mu_{N})$}      &  -1.33  & -1.31  & -1.31  & -1.30
  & -1.30 & -1.30  &-1.91 \\
   \centering{$g_{A}$}                  &  0.61   & 0.78   & 0.80   & 0.85
 & 0.86  &  0.89  &1.23  \\
 \end{tabular}
\label{table1}
\end{table}

\begin{table}
\caption{Numerical results obtained for some physical observables for some
different orders of truncation. The $4^{th}$ order corresponds to the standard
Skyrme Lagrangian.}
 \begin{tabular}{cdddddd}
    \centering{Order}  &  {4}  & {8}  & {12}  & {24}  & {60}  & {80} \\ \hline
    \centering{$f_{\pi}$ (MeV)}         &  64.6   & 73.6   & 75.4   & 76.8
 & 77.3  & 77.4 \\
    \centering{$10^{3} \epsilon^{2}$}   &  4.22   & 1.80   & 1.63   & 1.58
& 1.58  &      \\
    \centering{$<r^2_{E}>^{1/2}$ (fm)}  &  0.59   & 0.60   & 0.61   & 0.61
 & 0.61  & 0.61 \\
    \centering{$<r^2_{M}>^{1/2}$ (fm)}  &  0.92   & 0.88   & 0.87   & 0.86
 & 0.86  & 0.86  \\
   \centering{$\mu_{p} (\mu_{N})$}      &  1.87   & 1.89   & 1.89   & 1.89
& 1.89  & 1.89  \\
   \centering{$\mu_{n} (\mu_{N})$}      &  -1.33  & -1.32  & -1.31  & -1.31
  & -1.31 & -1.31 \\
   \centering{$g_{A}$}                  &  0.61   & 0.73   & 0.76   & 0.78
 & 0.79  &  0.79 \\
 \end{tabular}
\label{table2}
\end{table}

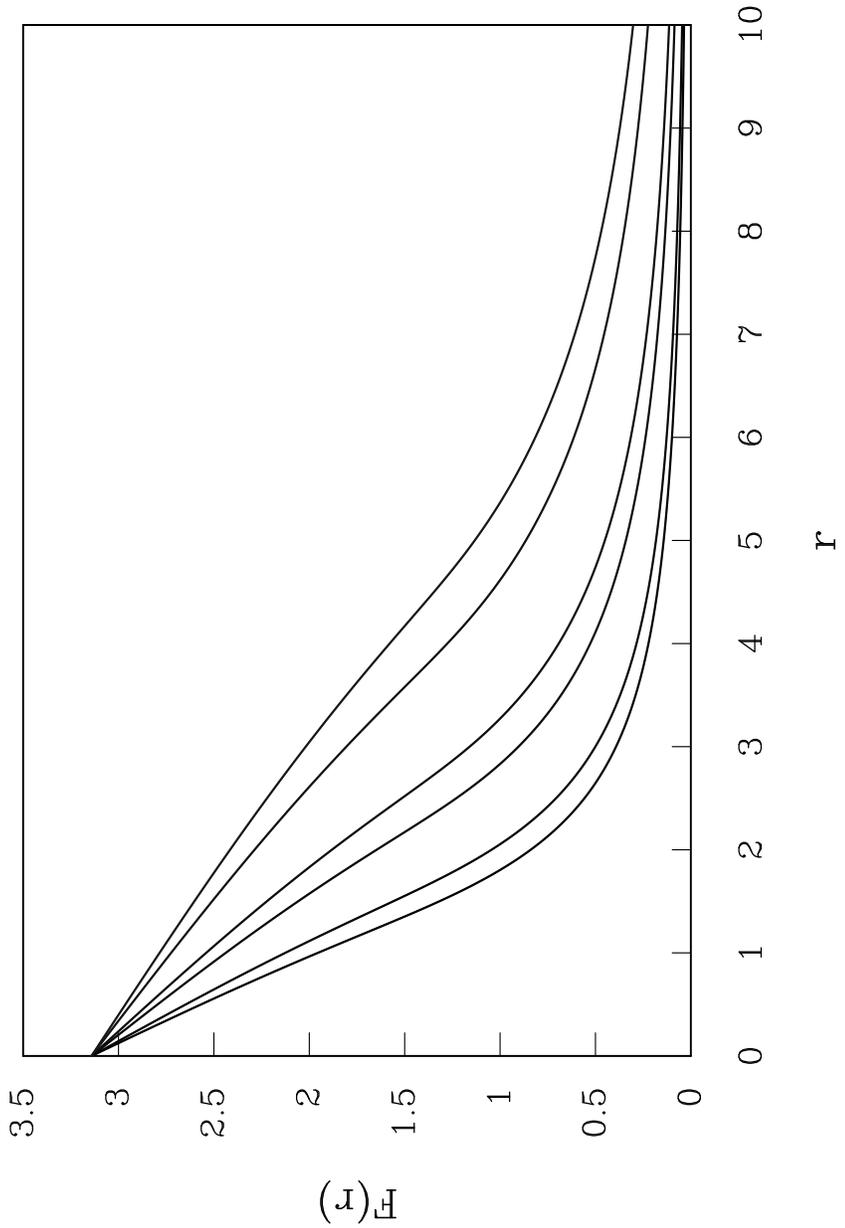
\begin{figure}
\caption{The chiral angle solution to (20) function of the radial distance
(in units of $a_0$) for values 1, 0.5, 0.1, 0.05, 0.01,
0.005 of the parameter $c$ defined in (18,19) at order 24. The lowest curve
in the graph corresponds to $c=1$ and the highest one to $c=0.005$.}
\label{fig1}
\end{figure}

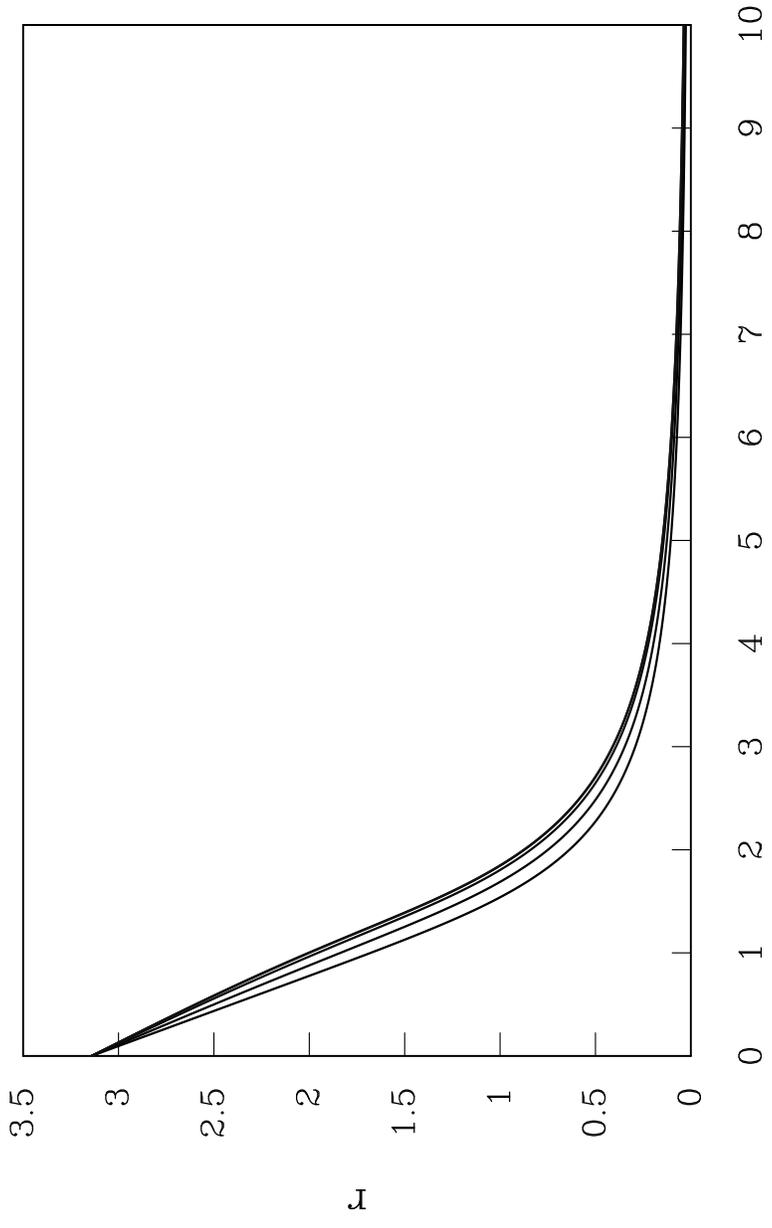
\begin{figure}
\caption{The chiral angle solution to (20) function of the radial distance
(in units of $a_0$) for orders 8, 12, 24, 40, 80 of truncation
with $c=1$. The lowest curve in the graph
corresponds to $8^{th}$ order truncation and the highest one
to $80^{th}$ order truncation.}
\label{fig1}
\end{figure}

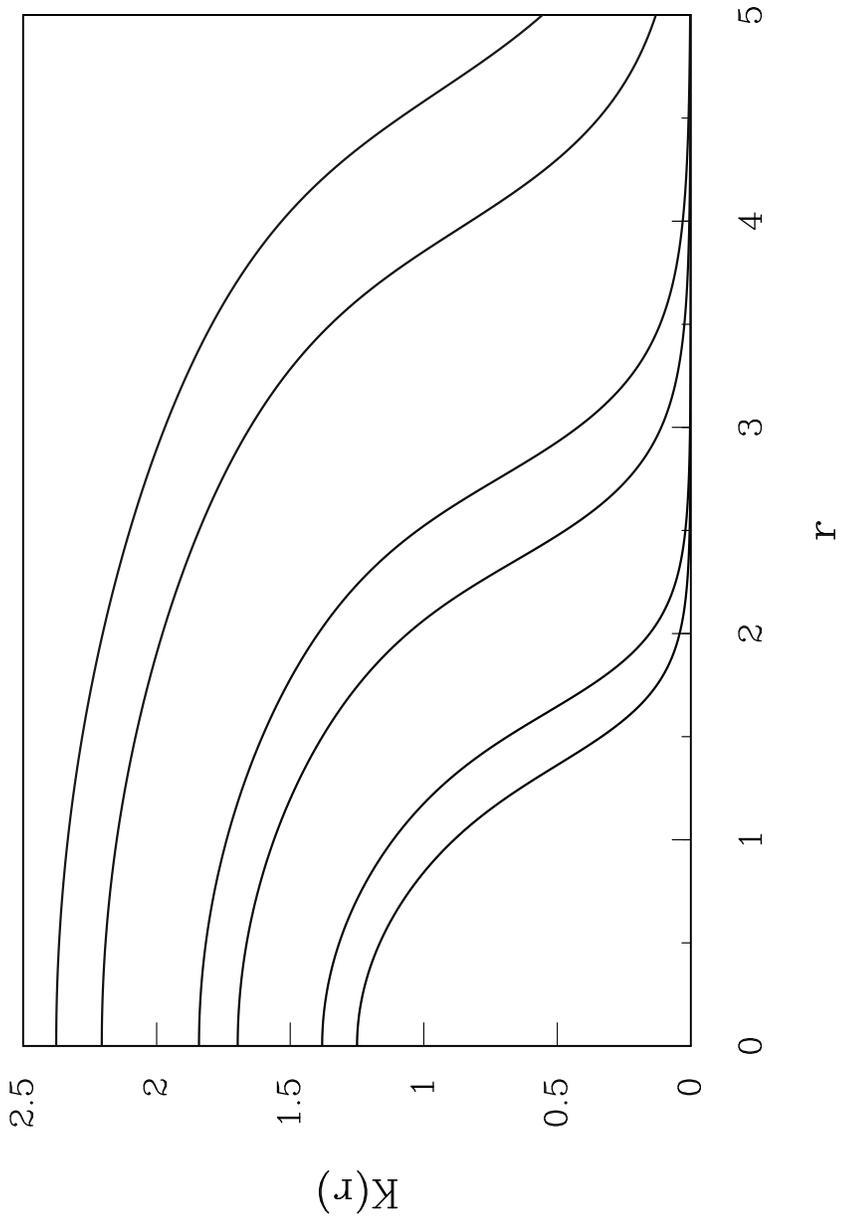
\begin{figure}
\caption{$K(r)$ defined in (19) function of the radial distance
(in units of $a_0$) for values 1, 0.5, 0.1, 0.05, 0.01,
0.005 of the parameter $c$ defined in (18,19) at order 24. The lowest curve
in the graph corresponds to $c=1$ and the highest one to $c=0.005$.}
\label{fig3}
\end{figure}

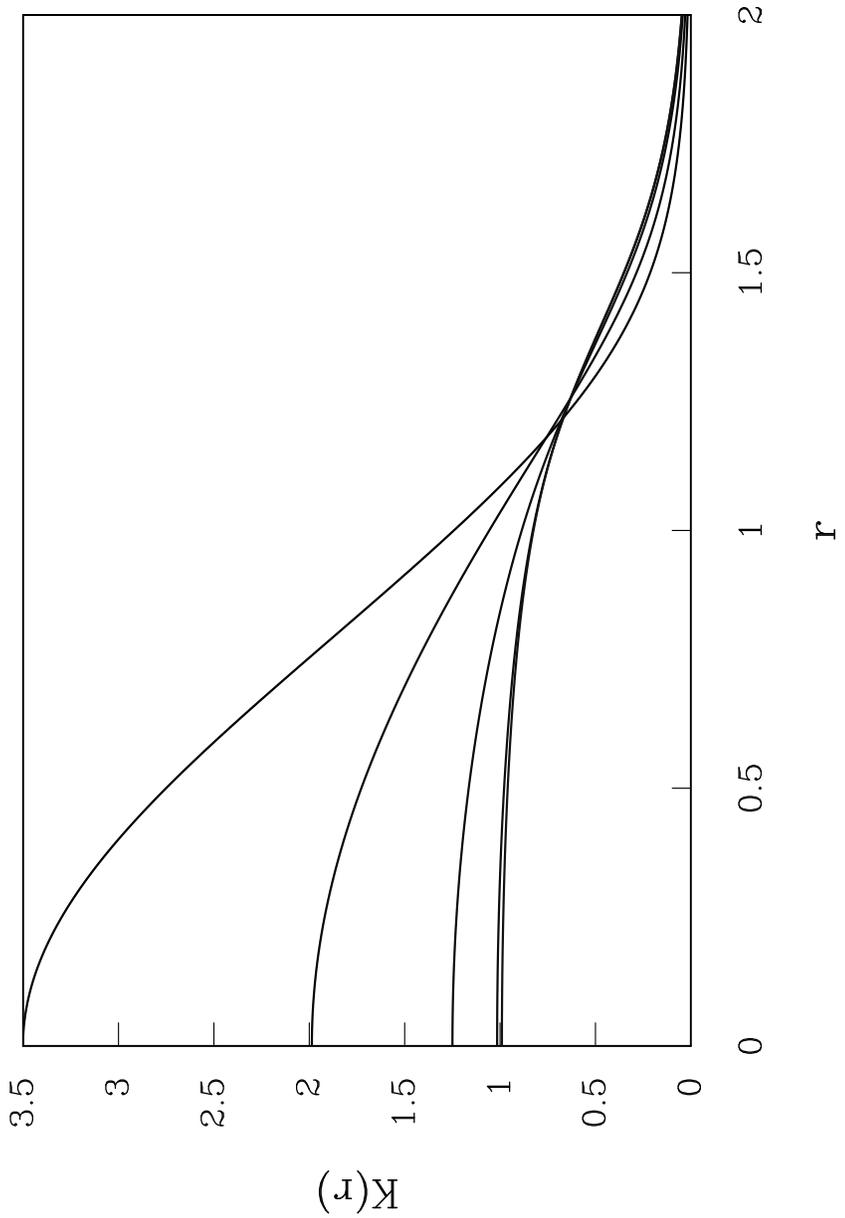
\begin{figure}
\caption{$K(r)$ defined in (19) function of the radial distance
(in units of $a_0$) for orders 8, 12, 24, 40, 80 of truncation
with $c=1$. The lowest curve in the graph
corresponds to $8^{th}$ order truncation and the highest one
to $80^{th}$ order truncation.}
\label{fig4}
\end{figure}

\end{document}